\documentclass{PoS}
\usepackage{graphicx}                            

\def\DU  {\mathop{{\cal D}\hbox{U}}}
\def\Dpsi {\mathop{{\cal D}\bar{\psi}{\cal D}\psi}}
\def\dd  {\mbox{d}}
\newcommand\detn[1]{\mbox{det}_{#1}}
\newcommand\tdetn[1]{\widetilde{\mbox{det}}_{#1}}
\def\Re {\mathop{\hbox{Re}}}
\title{New results using the canonical approach to finite density QCD}

\ShortTitle{New results using the canonical approach to finite
density QCD}

\author{\speaker{Anyi Li}, Andrei Alexandru, Keh-Fei Liu\\
        Department of Physics and Astronomy, University of Kentucky, Lexington KY 40506, USA\\
        E-mail: \email{anyili@pa.uky.edu}, \email{alexan@pa.uky.edu}, \email{liu@pa.uky.edu}}

\abstract{We present some new results regarding simulations of
finite density QCD based on a canonical approach. A previous study
has shown that such simulations are feasible, at least on small
lattices. In the current study, we investigate some of the issues
left open: we study the errors introduced by our approximation of
the action and we show how to tune it to reduce the cost of the
simulations while keeping the errors under control. To further
reduce the cost of the simulations, we check the reliability of
reweighting method with respect to the baryon number. Finally,
using these optimizations, we carry out the simulations at larger
densities than in our previous study to look for signals of a
phase transition.}

\FullConference{The XXV International Symposium on Lattice Field Theory\\
         July 30-4 August 2007\\
         Regensburg, Germany}

\begin{document}

\section{Introduction}
In recent years, full QCD simulations have become feasible due to
development of new algorithms and increasing computational power.
Lattice simulations using dynamical fermions can now be performed
at finite temperature and zero baryon density. However,
simulations at non-zero density remain a challenge for lattice QCD
due to the complex nature of the fermionic determinant where the
conventional Monte Carlo methods fail. The standard solution of
splitting the action into a real an positive part and a phase
fails due to the so called sign problem and overlap problem. A
method to address the overlap problem has been proposed
\cite{kfl05}. This method uses the canonical partition function as
a starting point, in contrast to the more common approach based on
the grand canonical partition function. Using this method,
simulations that employ an exact calculation of the determinant
\cite{afhl05} were carried out. It was shown that finite density
simulations based on the canonical partition function are feasible
and a program was outlined to look for the phase transition at
non-zero baryon densities. New simulations were carried out to
address some of the issues not covered in the previous study; in
this paper we report the results of these simulations.

Due to practical considerations, simulations based on the
canonical partition function have to use an approximation of the
action. The original study looked at deviations from the expected
quark number in the box to gauge the errors due to this
approximation. In this work, we use a more direct method: we
measure the same physical quantities while reducing the errors of
our approximation. We find that the errors are small except for
the chemical potential measured at the symmetric point of discrete
Fourier transform. As it turns out, the source of these errors is
very easy to understand and correct. This also leads us to a very
simple criterion that can be used to reduce the simulation cost
while keeping the errors at a minimum.

To further reduce the cost of our simulations we employ
reweighting. We analyze the reliability of reweighting in baryon
number using the methods we tested in \cite{lal06}. By comparing
the results of reweighting with the direct simulations we find
that we can extrapolate reliably by at least one baryon.

Finally, we use the ensembles generated for these studies to look
for a signal for phase transition at large baryon number. We
measure both the chemical potential and the plaquette distribution
and vary the baryon number while keeping the temperature fixed. We
scan the density space at two different temperatures close to the
critical temperature. We don't find any clear signal for a phase
transition on the lattices we studied.

\section{Canonical approach}

Finite density ensembles were generated using the canonical
approach~\cite{afhl05}. To build canonical partition function, we
start from the fugacity expansion of the grand canonical partition
function
\begin{equation} Z(V,T,\mu) = \sum_{k} Z_C(V, T, k) e^{\mu k/T},
\end{equation}
where $k$ is the net quark number and $Z_C$ is the canonical
partition function of the system. On the lattice, we can easily
compute the Fourier transform of the grand canonical partition
function
\begin{equation}
Z(V,T,\mu) = \int \DU \Dpsi e^{-S_g(U)- S_f(\mu; U,
\bar{\psi},\psi)}
\end{equation}
with imaginary chemical potential to get the canonical partition
function
\begin{equation}
Z_C(V, T, k) = \frac{1}{2\pi} \int_0^{2\pi} \mbox{d}\phi \,e^{-i k
\phi} Z(V, T, \mu)|_{\mu=i\phi T}.
\end{equation}
We will specialize this to the case of two degenerate flavors.
After integrating out the fermionic part, we get a simple
expression
\begin{equation}
Z_C(V, T, k) = \int \DU e^{-S_g(U)} \detn{k}
M^2(U)\label{eq:canonical},
\end{equation}
where
\begin{equation}
\detn{k} M^2(U) \equiv \frac{1}{2\pi}\int_0^{2\pi} \dd\phi\,e^{-i
k \phi} \det M(m, \mu;U)^2|_{\mu=i\phi T} ,
\end{equation}
is the  projected determinant with the fixed net quark number $k$.
$Z_C$ is the starting point for our simulations. For practical
reasons, we replace the continuous Fourier transition with its
discrete version
\begin{equation}
\tdetn{k} M^2(U) \equiv \frac{1}{N} \sum_{j=0}^{N-1} e^{-i k
\phi_j} \det M(U_{\phi_j})^2,~~~~~\phi_j=\frac{2\pi
j}{N}.\label{eq:fourier}
\end{equation}
The partition function that we use in our simulations is:
\begin{equation}
  \widetilde{Z}_C(V,T,k)\equiv\int DU e^{-S_g(U)}\Re\tdetn{k}M^2(U).
\end{equation}
For more details, we refer the reader to the original
paper~\cite{afhl05}.

We want to emphasize that the determinant of the fermionic matrix
needs to be calculated in every step while generating canonical
ensembles~\cite{kfl05}. It is well known that the calculation of
determinant calculation is very expensive even for a small lattice
(on a $4^4$ lattice, the dimension of the matrix is 3072). It is
also possible to simulate this action using a noisy estimator for
the determinant (Noisy Monte Carlo~\cite{jhl03}). In this paper,
we show results based only on the exact determinant calculation.

\section{Discretization errors and optimized Fourier transform}

To compute the Fourier transform of the determinant, we have to
evaluate the determinant for all possible phases. This is clearly
not feasible and we are forced to use a discrete version of the
Fourier transform. As shown in~\cite{afhl05}, the errors
introduced by this approximation are proportional to
$\exp(-[F(|N|-n)-F(n)]/T)$. It is easy to see that as $N$
increases the contamination decreases exponentially, while the
cost increases only linearly. We can then reduce the error for a
modest increase in computation time. The only problem is to
estimate the effect of the approximation. In~\cite{afhl05}, an
indirect method was proposed that gauged the effect of
contamination by measuring the difference between the expected
number of baryons in the box and the measured one.

In this study we use a direct approach: we increase $N$ and
measure the observables of interest again. One of them is the
absolute value of the Polyakov loop:
\begin{equation}
\left< |P| \right> = \frac{ \left< |P| \alpha \right>_0} {\left<
\alpha \right>_0},
\end{equation}
where
\begin{equation}
\alpha(U) = \frac{\Re\tdetn{k} M^2(U)}{\left|\Re \tdetn{k}
M^2(U)\right|},
\end{equation}
is the phase and $\left<\right>_o$ stands for the average over the
ensemble generated with measure $\left|\Re \tdetn{k}
M^2(U)\right|$. The other one is the baryon chemical potential
\begin{equation}
\label{baryon chemical potential}
\left<\mu\right>_{n_B}   =  \frac{F(n_B+1)-F(n_B)}{(n_B+1)-n_B} =
-\frac{1}{\beta}\ln \frac{\widetilde{Z}_C(3n_B+3)}{\widetilde{Z}_C(3n_B)}
= -\frac{1}{\beta}\ln \frac{\left< \gamma(U)\right>_o}{\left< \alpha(U)\right>_o}
\end{equation}
where
\begin{equation}
\gamma(U)\equiv  \frac{{\rm Re\,}\tdetn{3n_B+3} M^2(U)}
{\left|{\rm Re\,}\tdetn{3n_B} M^2(U)\right|}.
\end{equation}

In the equations above, $N$ appears implicitly in the definition
of the probability measure for the generated ensemble and in the
definition of the observables. The reason we use the same value in
these two definitions is that we want to have the generated
ensemble as close as possible to the target one. However, it is
possible to do reweighting by measuring the observables on a
target ensemble with $N=N_o$, on ensembles generated with $N=N_e$.
In other words, $N_e$ is used in defining the measure $\left| {\rm
Re\,}\tdetn{k}M^2\right|$ and in the definition of the
denominators for $\alpha(U)$ and $\gamma(U)$. $N_o$ is used in
defining the numerators for $\alpha(U)$ and $\gamma(U)$. The
expressions for the Polyakov loop and the chemical potential
remain the same in terms of $\alpha(U)$ and $\gamma(U)$.

The reason for this separation is that it is expensive to increase
$N_e$ since we have to evaluate the determinant $N_e$ times at
every accept-reject step; whereas, $N_o$ is relatively cheaper to
increase since we measure the projected determinant based on $N_o$
only on the saved configurations (a factor of 100 to 1000 times
less evaluations than in the previous case). On the other hand, we
do not want to have $N_o$ much larger than $N_e$ since this can
lead to an overlap problem.

To investigate the effects of our approximation we first increased
$N_o$ from $12$, the value used in our previous
study~\cite{afhl05}, to $24$. In this case, we could directly use
the ensembles already generated. In Table
\ref{poly_chem_direct_meas}, we present the results of this
comparison for $\beta=5.20$ on a $4^4$ lattice with $m_\pi\approx
1 {\rm GeV}$. We see that the Polyakov loop measurements are not
affected by this change, while the chemical potential measurements
at $k=3$ disagree. To understand this result we decided to carry
out a more direct check by generating an ensemble with $N_e=24$.

\newcommand\comment[1]{}
\comment{
\begin{table}[bh]
\centering
\begin{tabular}{|c|c|c||c|c|}
\hline
      &\multicolumn{2}{|c||}{Polyakov
      loop}&\multicolumn{2}{|c|}{Baryon chemical potential ($\mu_{n_B}/T$)}\\
\hline
      \textbf{\em k} & $N_o=12$ & $N_o=24$ & $N_o=12$ & $N_o=24$\\
\hline
      0 & 0.192(3)&0.192(3) & 3.40(21)  & 3.40(21)\\
\hline
      3 & 0.363(6)&0.363(6) & 3.09(11)  & 3.78(11)\\
\hline
\end{tabular}
\caption{Polyakov Loop and baryon chemical potential for $N_e=12$.
\label{polyakov chemical potential}}
\end{table}
}

\begin{table}[h]
\centering
\begin{tabular}{|c|c|c|c||c|c|c|}
\hline
      &\multicolumn{3}{|c||}{Polyakov
      loop}&\multicolumn{3}{|c|}{Baryon chemical potential ($\mu_{n_B}/T$)}\\
\hline
      \textbf{\em k} & $N_{e,o}=12$ & $N_e=12\,\,N_o=24$ & $N_{e,o}=24$ &
  $N_{e,o}=12$ & $N_e=12\,\,N_o=24$ & $N_{e,o}=24$\\
\hline
      0 & 0.192(3)& 0.192(3) & 0.193(8) & 3.40(21)  & 3.40(21) & 3.20(15) \\
\hline
      3 & 0.363(6)& 0.363(6) & 0.354(13) & 3.09(11)  & 3.78(11)  & 3.74(10) \\

\hline
\end{tabular}
\caption {Polyakov loop and baryon chemical potential: direct measurements
vs reweighting.
\label{poly_chem_direct_meas}}
\end{table}

In Table~\ref{poly_chem_direct_meas} the results from the direct
simulations are presented. We see that the Polyakov loop is the
same for all three cases. The baryon chemical potential for $k=3$
is very interesting: it agrees with the value for $N_e=12$ and
$N_o=24$ rather than the one produced with $N_e=N_o=12$. To
understand this, we first note that
\begin{equation}
\label{eq:partition}
  \widetilde{Z}_C(V,T,k)= \sum_{m=-\infty}^{m=+\infty}
Z_C(V,T,k+mN),
\end{equation}
where $Z_C(V,T,k+mN)$ is defined using the continuous Fourier
transform. Using the above property together with the fact that
$Z_C(V,T,k)= Z_C(V,T,-k)$, we get for the symmetric point ($k=6$
for $N=12$)
\begin{eqnarray}
\label{eq:expansion}
  \widetilde{Z}_C(V,T,6)&=& \sum_{m=-\infty}^{m=+\infty}
Z_C(V,T,6+m12) \nonumber \\
&=&\ldots + Z_C(V,T,-6)+ Z_C(V,T,6)+ Z_C(V,T,18)+\ldots \nonumber \\
&\approx& Z_C(V,T,-6)+ Z_C(V,T,6) = 2Z_C(V,T,6).
\end{eqnarray}
Using Eq. (~\ref{baryon chemical potential}) we get
\begin{equation}
\left<\mu\right>_{k=3} ^{N=12}/T = -\ln \frac{\tilde{Z}_C(V,T,6)}{\tilde{Z}_C(V,T,3)}
\approx -\ln \frac{2 Z_C(V,T,6)}{Z_C(V,T,3)} =
\left<\mu\right>_{k=3} ^{N=\infty}/T - \ln 2
\end{equation}
The $\ln2$ correction changes $3.09(11)$ in Table
\ref{poly_chem_direct_meas} to $3.78(11)$ which agrees with the
result of reweighting and direct simulation.

Although the baryon chemical potential measurement at the
symmetric point could be corrected by an extra $\ln 2$, the
contamination from the other sector can introduce errors in other
observables. To be on the safe side it is better to carry out
simulations with $N_e$ larger than twice the value of $k$. Since
the contamination errors seem much smaller than the statistical
error we will carry out our simulations with $N_e = 2k+3$. Before
we move on, we want to point out that these conclusions might
change when we lower the mass of the quark as pointed out
in~\cite{afhl05} the contamination is suppressed by a factor
proportional to the baryon free energy which is expected to
decrease as we lower the quark mass.

\section{Reliable range of reweighting}

\begin{figure}[th]
  \centering
  \includegraphics[scale=0.83,clip=true]{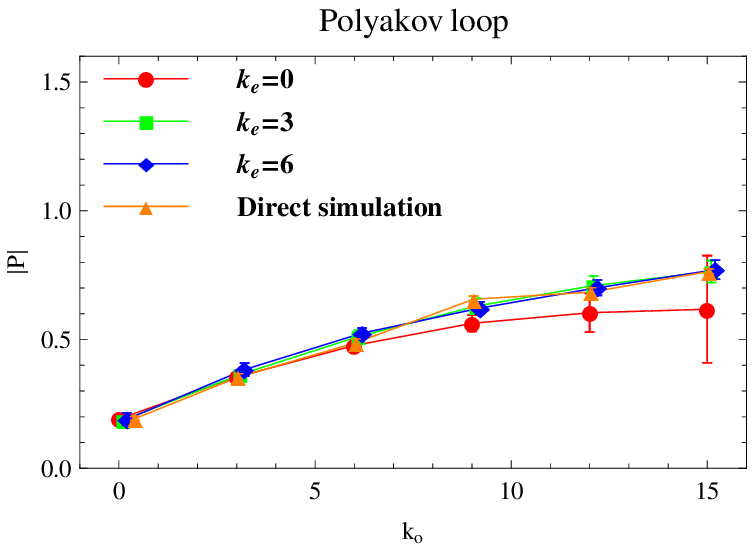}
  \includegraphics[scale=0.8,clip=true]{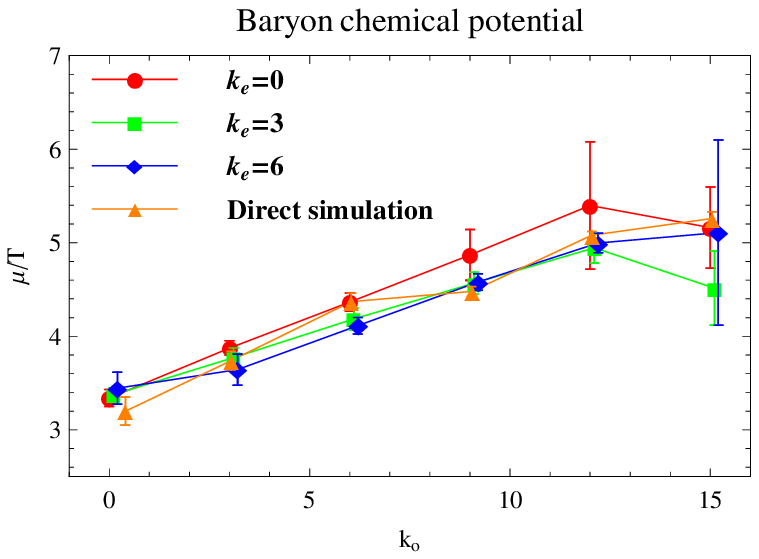}
  \caption{Left: Polyakov loop from reweighting against direct simulations.
  Right: The same for baryon chemical potential. $k_o=k_e$ except for the case of reweighting. \label{reweighting}}
\end{figure}

In the previous section, we have shown that we can generate
ensembles based on a smaller $N_e$ and use a form of reweighting
to get results corresponding to a different ensemble. We can
generalize this reweighting to change not only $N$ to $N_o$ but
also $k$ from the value used to generate the ensemble, $k_e$, to
the one that will be used in the observable, $k_o$. The only
difference is that one needs to adjust the phase factors
accordingly. To determine the reliability of the method, we
compare the results of reweighting to the ones derived from the
direct measurements. In Fig.~\ref{reweighting}, we present the
results based on 3 ensembles with $k_e=0$, $3$ and $6$. They are
all generated at $\beta=5.20$, using $N_e=12$ and the same quark
mass. We use the reweighting to produce results corresponding to
$N_o=48$ and $k_o=0$, $3$, $6$, ..., $15$ and we compare these
results with those of direct measurements. For the direct
measurements we use $k_e=k_o$ and $N_e=N_o$ and we set $N_e=24$
for $k_e=0$, $3$ and $6$, $N_e=27$ for $k_e=9$ and $N_e=39$ for
$k_e=12$ and $15$.

We find that reweighting is reliable for quite a large range of
baryon numbers. The fact that we can extrapolate from $k_e=0$ to
$k_o=9$ is quite surprising but we cannot thrust this property to
remain true at different temperatures and volumes. However, our
plan was to use reweighting to interpolate rather than
extrapolate: we plan to scan the baryon number space sparsely
(skipping some values of $k$) and filling these gaps by
reweighting. This study suggests that this method should produce
reliable results if $|k_e-k_o|$ is no larger than 6.

\section{Possible phase diagram}

We now turn to the physical problem -- the QCD phase diagram. It
is expected that two flavor QCD phase diagram will have a critical
point at non-zero chemical potential where the first order phase
transition turns to a crossover. In the canonical ensemble, the
first order phase transition will be represented by a coexistence
region. To search for the phase boundaries we scan the phase
diagram by varying the density while keeping the temperature
fixed. The baryon chemical potential should exhibit an
``S-shape''~\cite{fk06} as one crosses the coexistence region. The
Maxwell construction can then be used to determine the phase
boundaries. Another possible signal can be detected by looking at
the histogram of the plaquette.~\cite{se07}

\begin{figure}[th]
  \centering
  \includegraphics[scale=0.8,clip=true]{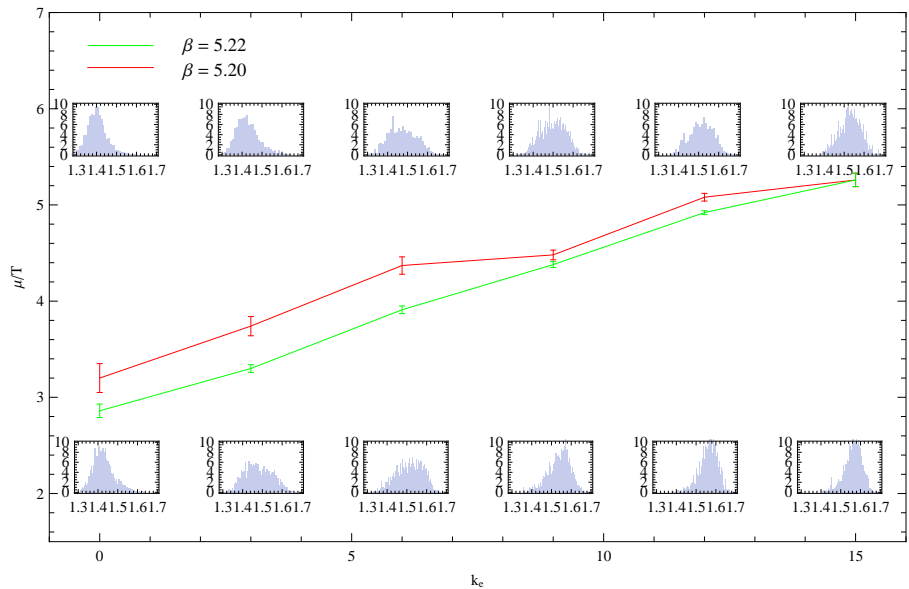}
  \caption{Chemical potential and plaquette histograms for temperatures
close to the transition temperature.
\label{hist_chem}}
\end{figure}

To test these ideas, we have generated two sets of ensembles at
two temperatures close to the critical temperature. In
Fig.~\ref{hist_chem}, we plot the baryon chemical potential and
the histogram of the plaquette.
We notice a peak shift in the histogram plot but the chemical
potential does not exhibit a clear ``S-shape''. Although there is
no evidence of an ``S-shape'', the peak shift suggests that a
transition happens between 3.6 to 18 times the normal nuclear
matter density whereas Ph. de Forcrand and S. Kratochvila found
the range to be 1 to 10 times of the nuclear density~\cite{fk06}.
The difference may be due to the heavy quark mass we use in our
study and the fact that they studied a 4 flavor QCD. Our heavy
quark mass may also have suppressed the fluctuations and hence
obscured the ``S-shape''.

\section{Summary}

In this paper, we present new results that address some of the
issues not covered in our previous study. We show that the errors
introduced by using a discrete Fourier transform are smaller that
the statistical errors with one exception that we analyze and show
how to correct. We find that the optimal strategy is to set
$N_e=2k+3$ to reduce the simulation cost while keeping the errors
under control. We also analyze the reliability of reweighting in
baryon number by comparing the results of reweighting with direct
simulations. We find that the method is reliable for a larger
range than we expected and this will allow us to further reduce
the cost of our simulations in the future. Finally, we have used
the ensembles generated for this study to look for the phase
transition at larger baryon numbers. Although the chemical
potential plot does not show a clear signal for phase transition,
the peak shift in the plaquette histogram from small densities to
large densities suggests a transition.

We plan to continue this study with simulations on a larger
lattice ($6^3\times4$) where we hope to be able to detect a phase
transition. We also plan to use improved gauge and fermionic
actions to run simulations at smaller quark masses.


\begin{thebibliography}{99}
  \bibitem{kfl05} K.~F.~Liu,  \emph{Edinburgh 2003, QCD and Numerical Analysis Vol. III} (Springer, New York, 2005) 101 [{\tt
  arXiv:hep-lat/0312027}].
  \bibitem{afhl05} A.~Alexandru, M.~Faber, I.~Horváth, K.~F.~Liu, \emph{Phys. Rev.} \emph{D\bf72} (2005) 114513
   [{\tt arXiv:hep-lat/0507020}].
  \bibitem{lal06} A.~Li, A.~Alexandru, K.~F.~Liu, \pos{PoS(LAT2006)030} [{\tt arXiv:hep-lat/0612011}]
  \bibitem{jhl03} B.~Joo, I.~Horváth, and K.~F.~Liu, \emph{Phys. Rev.} \emph{D\bf67} (2003) 074505
  [{\tt arXiv:hep-lat/0112033}]. A.~Alexandru {\sl et al.}, \pos{PoS(LATTICE
  2007)167}.
  \bibitem{fk06} Ph.~de~Forcrand, S.~Kratochvila, \emph{Nucl. Phys.} \emph{B} (Proc. Suppl.) {\bf 153} (2006) 62
  \bibitem{se07} S.~Ejiri [{\tt arXiv:hep-lat/0706.3549}]. S.~Ejiri, \pos{PoS(LATTICE 2007)181} [{\tt
  arXiv:hep-lat/0710.0653}].

\end{thebibliography}
\end{document}